\documentclass[pra,twocolumn]{revtex4}
\usepackage{amsmath}
\usepackage{amsfonts}
\usepackage{amssymb}
\usepackage{graphics}
\usepackage{graphicx}

\def\sbra#1{\langle #1|} 
\def\sket#1{|#1\rangle}

\begin{document}
\title{Particle-number scaling of the phase sensitivity in realistic Bayesian twin-mode Heisenberg-limited interferometry}
\author{Raphael Pooser}
\author{Olivier Pfister}
\email[Corresponding author. ]{opfister@virginia.edu}
\affiliation{Department of Physics, University of Virginia, 382 McCormick Road, Charlottesville, VA 22904-4714, USA}

\begin{abstract}
We investigate the scaling of the phase sensitivity of a nonideal Heisenberg-limited interferometer with the particle number $N$, in the case of the Bayesian detection procedure proposed by Holland and Burnett [\prl {\bf 71}, 1355 (1993)] for twin boson input modes. Using Monte Carlo simulations for up to $10\,000$ bosons, we show that the phase error of a nonideal interferometer scales with the Heisenberg limit if the losses are of the order of $N^{-1}$. Greater losses degrade the scaling which is then in $N^{-1/2}$, like the shot-noise limit, yet the sensitivity stays sub-shot-noise as long as photon correlations are present. These results give the actual limits of Bayesian detection for twin-mode interferometry and prove that it is an experimentally feasible scheme, contrary to what is implied by the coincidence-detection analysis of Kim {\em et al.}\ [\pra {\bf 60}, 708 (1999)].
\end{abstract}
\pacs{
}
\maketitle
\section{Introduction}
\begin{figure}[htb]
\begin{center}
\begin{tabular}{c}
\includegraphics[height= 1in]{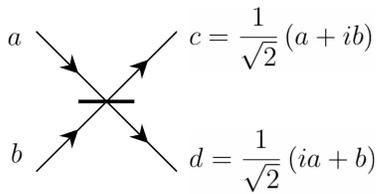}
\end{tabular}
\end{center}
\caption{Beam splitter. Reflection/transmission coefficients: $r=-i t=2^{-1/2}$. $a,b,c,d$: annihilation operators. The wave fronts overlap perfectly, i.e.\ the beams are aligned so that their wave vectors verify $\hat k_{c,d}=\hat k_{b,a}$.}
\label{f:bs}
\end{figure}

Heisenberg-limited quantum interferometry (HLI) is of interest for ultra-precise measurement of phase and energy differences \cite{hli}. Possible applications include the enhancement of atomic frequency standards and of gravitational-wave interferometers. Although optical interferometry often comes to mind first, interferometry at the Heisenberg limit concerns any bosonic quantum wave, and even fermionic ones \cite{yurke_f}. The absolute minimum error on the phase difference $\phi_-$ between the two interferometer arms, for $N$ detected bosons, is the Heisenberg limit (HL),
\begin{equation}
\Delta\phi_- \sim \frac{1}{N},	\label{e:hl}
\end{equation}
to be compared to the shot-noise limit of the beam splitter (SNL),
\begin{equation}
\Delta\phi_- \sim \frac{1}{\sqrt{N}}. \label{e:snl}
\end{equation} 
These limits arise from the quantum interference of the two input modes, $a$ and $b$, on the first beam splitter (boson-mode splitter) of the interferometer (Fig.~\ref{f:bs}).
This interference defines the quantum noise $\Delta N_-=N_a-N_b$ of the photon number difference at the output of the beam splitter (i.e.\ inside the interferometer). The quantum noise $\Delta \phi_-=\phi_a-\phi_b$ of the phase difference is then given by the number-phase Heisenberg inequality
\begin{equation}\label{e:hi}
\Delta N_-\Delta\phi_- \geq 1.
\end{equation}
In conventional interferometry, one of the inputs, say $b$, is a vacuum mode and the result of the interference is a binomial probability law, which yields the SNL, whatever the state of the bright input, $a$, is \cite{caves80}.

In order to achieve HLI, it is necessary to modify the vacuum input $b$. This can be done by squeezing a vacuum-field quadrature, as first proposed by Caves \cite{caves81} and experimentally realized by Xiao {\em et al.} \cite{xiao} and Grangier {\em et al.} \cite{grangier}. Numerous other schemes have been proposed since \cite{hli}, in particular by Yurke, McCall, and Klauder \cite{yurke} and Holland and Burnett \cite{holland}.  

This paper focuses on the latter proposal, whose experimental implementation is in progress in our group \cite{feng,feng2,feng3}. It is based, for reaching the HL, on the use of indistinguishable twin Fock states
\begin{equation}
\sket{\hat k_a,\omega,\hat\epsilon;n}\otimes \sket{\hat k_b,\omega,\hat\epsilon;n},
\label{e:twinFock}
\end{equation}
where $\hat k$ is the unit wave vector, $\omega$ the frequency, $\hat\epsilon$ the unit polarization vector, and $n$ the photon (boson) number. It is remarkable that, in the same manner as for the SNL \cite{caves80}, common-mode quantum statistics do not matter and the HL can be reached as efficiently by using superpositions or mixtures of twin Fock states \cite{kim}, or twin beams, whose density matrix is of the form  
\begin{equation}
\rho = \sum_{n,n'} \rho_{nn'} \sket{\hat k_a; n}\sbra{\hat k_a; n'}\otimes\sket{\hat k_b; n}\sbra{\hat k_b; n'},
\label{e:twinbeam}
\end{equation}
with all quantum numbers identical between modes $a$ and $b$, except the wave vectors --- which will be exactly matched at the beam splitter (Fig.~\ref{f:bs}). 

Equations (\ref{e:twinFock},\ref{e:twinbeam}) describe perfectly correlated interferometer inputs. The physics of the HL in this case lies solely in the generalized Hong-Ou-Mandel interference \cite{hom} of the twin Fock states (\ref{e:twinFock}) at a beam splitter. Superpositions or mixtures (\ref{e:twinbeam}) do not alter the result \cite{kim}. 

This has been demonstrated experimentally using a pair of independently number-squeezed picosecond-pulsed beams from optical fibers \cite{silb} and, more recently, by our group with ultrastable milliwatt-level twin beams emitted by a type II OPO electronically phase-locked at frequency degeneracy \cite{feng3}. 

Good approximations of twin boson modes have also been implemented using internal atomic states of  trapped ions \cite{meyer} and other proposals include internal atomic states of Bose-Einstein condensates \cite{orzel,bouyer} and orthogonal phonons in a single trapped ion \cite{djw}. In the following, we will consider the case of photons, but the reader should bear in mind that the generalization to bosons is mathematically immediate. Some physical analogs of the optical beam splitter are well known: for example, a $\pi/2$ resonant laser pulse coupling two atomic energy states is isomorphic to a balanced beam splitter overlapping two optical paths. Both are SU(2) systems and can be described by the Schwinger representation \cite{schwinger}.

A particularity of the Holland-Burnett scheme lies in its nonconventional detection process. When a twin mode input is used, the expectation values for the output photon numbers of an interferometer are independent of the phase difference $\phi_-$ \cite{holland,bouyer,kim}. Direct detection of the average output intensities, i.e.\ of the second-order moment of the input field, cannot therefore be used. The fourth-order moment is phase dependent, as has been noted first by Bouyer and Kasevich \cite{bouyer}, however, a simple derivation of the variance of that measurement shows that its signal-to-noise ratio is very low: $\sqrt 2$ at best \cite{kim}, which lowers detection accuracy. 

Coincidence detection can dramatically improve the signal-to-noise ratio of a fourth-order moment measurement and a proof-of-principle experimental demonstration has been carried out for $N=2$ \cite{kuzmich}. It is, however, not clear how to extend coincidence measurements to photon numbers $N>2$. What's more, an analysis of coincidence measurements in the presence of detector losses has shown extremely unfavorable results: in particular, it seems that the losses need to be smaller than $N^{-1}$ for this approach to yield even sub-SNL results \cite{kim99}. These unfavorable results have cast a doubt on the experimental feasibility of twin-mode HLI \cite{burnett}.

However, we would like to establish here that such is not the case if one considers the original measurement strategy proposed by Holland and Burnett, which does not rely on coincidence detection but on a Bayesian reconstruction procedure from dynamical photocurrent measurements \cite{holland}. The present paper thus focuses on Bayesian twin-mode interferometry in the presence of detection losses, and on its scaling with the photon number.

The influence of losses on Bayesian HLI was investigated by numerical simulation in a previous paper \cite{kim}, but the results were limited to low photon numbers (100) and did not address the question of the scaling of the sensitivity with $N$. In this paper, we carry out numerical simulations of Bayesian measurements for photon numbers ranging from $N=10$ to $10\,000$, in the presence of arbitrary detector losses. The results are very different from the ones obtained for coincidence detection and the prospects for experimental realizations are not as unfavorable as one might have feared from the conclusions of \cite{kim99}.

In Section II of this paper, we briefly review quantum interferometry and the Bayesian measurement procedure. We then present our numerical simulations in Section III and conclude.   

\section{Twin-mode HLI}

\begin{figure}[htb]
\begin{center}
\begin{tabular}{c}
\includegraphics[scale=1]{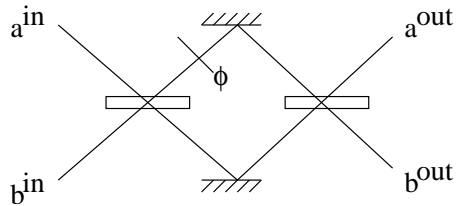}
\end{tabular}
\end{center}
\caption{Mach-Zehnder interferometer.} \label{f:mz}
\end{figure}

All optical interferometers can be modeled using a Mach-Zehnder model (Fig.~\ref{f:mz}), 
i.e.\ a first beam splitter ``splits" the input light beam into two paths $a$ and $b$ which experience a phase difference to be measured and the fields are then ``recombined" by a second beam splitter. The goal is to measure this phase difference with high precision.  As mentioned earlier, the output statistics of the first beam splitter determine the sensitivity and noise properties of the interferometer. 

For conventional interferometry, the input state
\begin{equation}
\sket{\hat k_a,\omega,\hat\epsilon;N}\otimes \sket{\hat k_b,\omega,\hat\epsilon;0},
\label{e:singleFock}
\end{equation}
yields a binomial law for the beam splitter output statistics $\Delta N_-\propto\sqrt N$ \cite{caves80}. If the input is a minimum uncertainty state --- equality in (\ref{e:hi}) --- the standard deviation of the phase difference is then the SNL (\ref{e:snl}). 

A squeezed-vacuum input state can be used too: 
\begin{equation}
\sket{\hat k_a,\omega,\hat\epsilon;\alpha}\otimes \sket{\hat k_b,\omega,\hat\epsilon;0,r},
\label{e:vacsq}
\end{equation}
where $r$ is the squeezing parameter. If the squeezing is in quadrature with the coherent state $\sket{\alpha}$, phase-difference squeezing is obtained at the output of the beam splitter \cite{caves81,wm}. 

We now turn to twin-mode Heisenberg-limited interferometry, where the input state is that of Eq.~(\ref{e:twinFock}). The detail of the theory is given in \cite{holland,kim} and briefly recalled here. As mentioned in the Introduction, the expectation value of the output intensity of a Mach-Zehnder interferometer is independent of the phase shift  $\phi$ between the two arms.
The Bayesian detection method is therefore used. It consists of burst measurements of the output intensities, rather than average measurements yielding the expectation value. The probability distribution of the output intensity is easy to derive \cite{kim} in the Schwinger representation \cite{schwinger} and is given by
\begin{equation}
P(j,m,\phi) =  \frac{(j-m)!}{(j+m)!}[P^{m}_{j}(\cos\phi)]^{2},
\label{e:probdist}
\end{equation}
where $j=(n_a+n_b)/2$ and $m=(n_a-n_b)/2$, $n_{a,b}$ are the measured photon numbers at each output, and where $P_j^m(\cos\phi)$ is an associated Legendre polynomial. Note that, if there are no losses, $j=N$ always for input state (\ref{e:twinFock}) because of unitarity. 

\begin{figure}[htb]
\begin{center}
\begin{tabular}{c}
\includegraphics[scale=0.6]{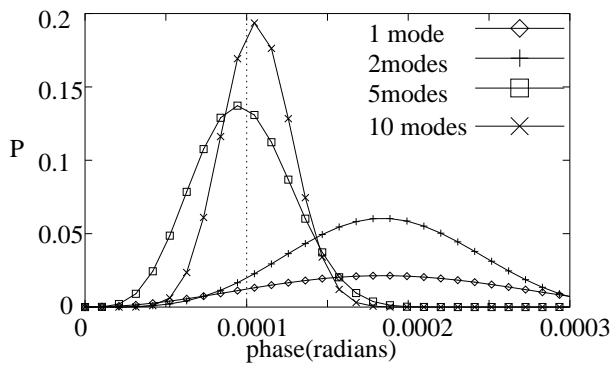}
\end{tabular}
\end{center}
\caption{Simulated Bayesian twin-mode phase probability distribution for a phase of 0.0001 radians (denoted by the vertical line) and several burst measurements (``modes") of 10000 photons each, at 100\% efficiency.} \label{f:graph_modes10000_100}
\end{figure}

Assuming each measurement result independent of one another, the probability of a set of $k$ burst measurements $[(j_1,m_1);\dots; (j_k,m_k)]$ is simply the product
\begin{equation}
P^{(k)}(\phi) = \prod_{i=1}^k P(j_i,m_i,\phi).
\end{equation}
When plotted against $\phi$ (Fig.~\ref{f:graph_modes10000_100}), $P^{(k)}$ is fairly broad at $k=1$ but narrows down quickly to a sharp peak with $\Delta\phi\sim N^{-1}$ at $k = 10$. 
In the presence of optical losses, analytical treatment is very complicated and numerical simulation is necessary, in particular in order to elucidate the scaling of the phase error with the photon number. This is the subject of the next section.
 
\section{Numerical simulations of nonideal Bayesian twin-mode HLI}

\subsection{Method}

The monte carlo simulation for this study uses twin Fock states as the input.  The detectors along with their efficiencies are simulated on the output ports.  For each phase difference measurement within a specified range, the program samples a number of one-shot instantaneous intensity difference measurements at the photodetectors.  This corresponds to a number difference $m$ between the output beams.  While the average number difference is expected to be zero over long measurement times, instantaneous fluctuations are not.  Thus in any given measurement the number difference may be drastically different from zero.  Using this difference the program reconstructs a probability distribution for the phase, for each burst measurement of the number difference $m$.  

Each measurement gives a different $m$, most of the time, and can also give a different $j$ if detectors are not 100\% efficient or the input state is not perfectly correlated. Previous work showed that both cases are equivalent, even though the nonideal-detector case is much easier to handle numerically, using Legendre in lieu of general Jacobi polynomials \cite{kim}. We therefore claim that a general analysis can be carried out by solely accounting for detection losses and assuming ideal initial correlations. 

The ``measured" peak of Fig.~\ref{f:graph_modes10000_100} is centered on the ``true" phase value, fixed by the program, and the peak width gives the measurement uncertainty.  The code calculates this uncertainty and shows a diminishing value as the number of bursts in the measurement increases.  This occurs even if the peak predicted by the probability distribution is not in fact centered at the proper phase value (postulated by the program before measuring the intensity difference).  In such a situation, the distance of the measured peak from the real phase value becomes the dominant factor in the uncertainty, as the width of the distribution is generally much smaller than this value.  On average this distance is much smaller than the SNL for all types of quantum efficiencies studied, and it reaches the HL for 100\% efficiency.

\begin{figure}[htb]
\begin{center}
\begin{tabular}{c}
\includegraphics[scale=0.6]{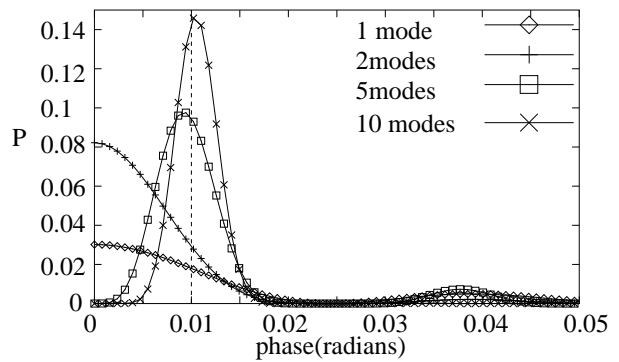}
\end{tabular}
\end{center}
\caption{Simulated Bayesian twin-mode phase probability distribution for a phase of 0.01 radians (denoted by the vertical line) and several burst measurements (``modes") of 100 photons each, at 99\% efficiency.} \label{f:graph_modes100_99}
\end{figure} 

Several total input photon numbers were used in the program, $10$, $100$, $1000$, and $10000$ at efficiencies of $90\%$, $99\%$, $99.9\%$,  $99.99\%$, and $100\%$.

For each phase difference specified there is an associated Legendre polynomial evaluated that is used to simulate the detected total photon number and number difference between the detectors.  The losses are computed by operating on the total number of photons in each beam as they exit the second beam splitter using a binomial distribution weighted by the detector efficiency.  The total number of photons on the output ports is simulated using a random possibility to transmit or reflect each photon at the beam splitters, which depends on the phase.

For examples of the probability distributions, or synonymously, phase distributions, see Figs.~\ref{f:graph_modes10000_100} and \ref{f:graph_modes100_99} (note the vertical axes are in arbitrary units that indicate relative probability compared to points on the same phase distribution).

\subsection{Results}

All of the figures in this section refer to uncertainties on measurements of phase angles $\phi=1/N$.  This means the data points in each graph refer to the uncertainty in a measurement for a different phase angle from one data point to the next. The most important results are found for phase differences ranging from zero to a few times $1/N$. There is no ``optimal" phase angle in this range, unlike what Kim et al.\ found for coincidence detection \cite{kim99}. For large phase angles ($\phi \rightarrow \pi/2$), one cannot acheive better than the SNL.  Further, the uncertainty is found to be quasi-ideal for nonideal quantum detection efficiencies $\eta=\eta_N$ such that
\begin{equation}
\eta_N = 1 - \frac{1}{N}.	\label{e:eff}
\end {equation}
We have also found the remarkable result that the phase measurement uncertainty scales as $1/N$ for such nonideal quantum efficiencies, as in the case of 100\% efficient detectors. This defines a margin of experimental conditions within which the phase uncertainty may be further increased below the SNL by merely increasing the boson number $N$, but not the efficiency itself (or the effective photon correlation). Even though $\eta_N$ may be thought of impossible to reach at optical powers, such may not be the case with matter waves, where very high efficiency detection is possible.

While detector efficiencies smaller than $\eta_N$ do not yield such a favorable scaling of the phase measurement error ($1/\sqrt N$ instead of $1/N$, like squeezing measurements \cite{wm}), it is nevertheless found that the Bayesian measurement method can always give a better precision than the SNL, which, again, was not the case for coincidence detection \cite{kim99}. In this regime, the gain in phase error is solely conditioned by the improvement of the effective correlation, as is the case for vacuum squeezing experiments \cite{wm}.

\begin{figure}[htb]
\begin{center}
\begin{tabular}{c}
\includegraphics[width= 3in]{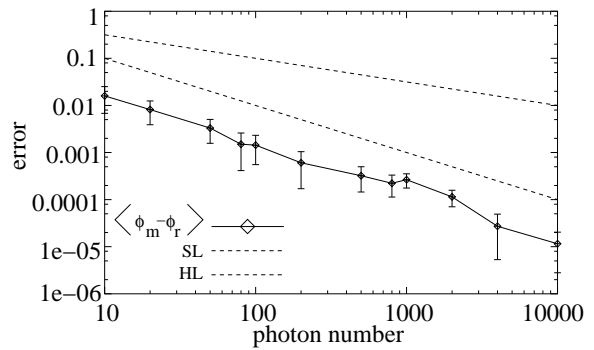}
\end{tabular}
\end{center}
\caption{Average of measured minus real phase for 100\% efficiency (log-log scale).} \label{f:graph100}
\end{figure}

We now turn to the simulation plots. For 100\% efficiency (Fig.\ \ref{f:graph100}) the phase error follows $1/N$ as expected and one can also see that both the simulated SNL and HL are slightly below $1/\sqrt{N}$ or $1/N$. 
We attribute this to the averaging process of multiple-burst measurements.  In situations involving linear averaging, one gets a reduction factor of the noise of $1/\sqrt{p}$, where $p$ is the number of data samples.  In our case, the measurement probabilities are multiplied together, not averaged. Nevertheless, the results are found to lay within less than an order of magnitude of a line consistent with the HL reduced by $1/\sqrt{p}$, where $p$ is the number of probability distributions, or bursts. 

\begin{figure}[htb]
\begin{center}
\begin{tabular}{c}
\includegraphics[width= 3in]{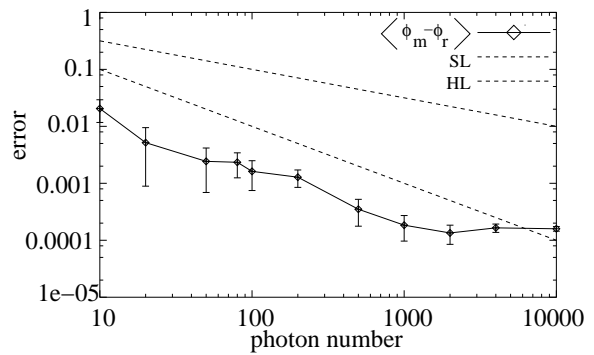}
\end{tabular}
\end{center}
\caption{Average of measured minus real phase for 99.9\% efficiency (log-log scale).} \label{f:graph999}
\end{figure}

\begin{figure}[htb]
\begin{center}
\begin{tabular}{c}
\includegraphics[width= 3in]{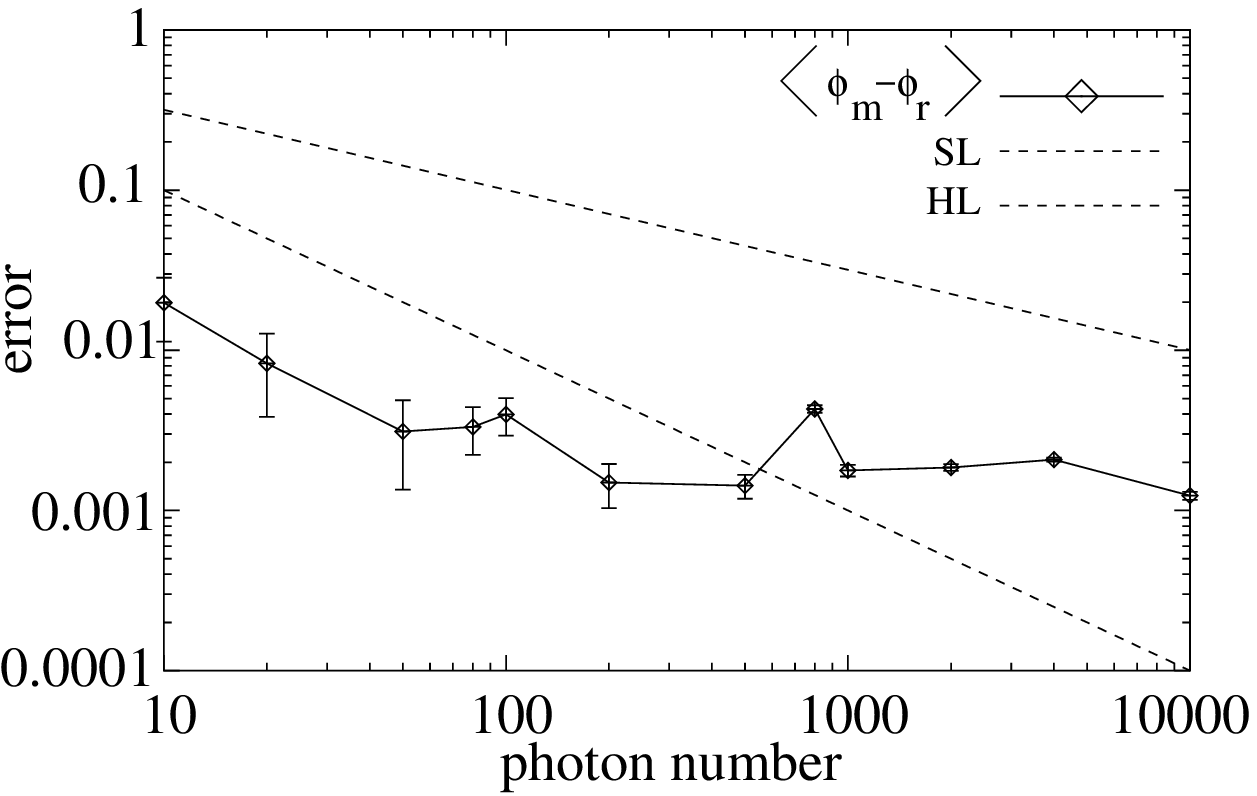}
\end{tabular}
\end{center}
\caption{Average of measured minus real phase for 99\% efficiency (log-log scale).} \label{f:graph99}
\end{figure}

\begin{figure}[htb]
\begin{center}
\begin{tabular}{c}
\includegraphics[width= 3in]{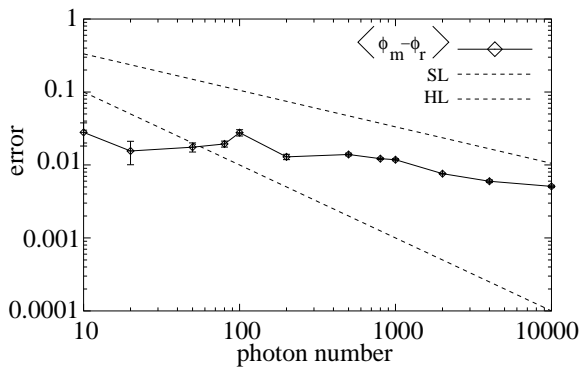}
\end{tabular}
\end{center}
\caption{Average of measured minus real phase for 90\% efficiency.} \label{f:graph90}
\end{figure}

For nonideal efficiencies $\eta<100\%$ (Figs.\ \ref{f:graph999} to \ref{f:graph90}), we find that the phase error follows $1/N$ for low photon numbers before crossing the HL line to follow the SNL at $1/\sqrt N$. Our simulations indicate that the inflexion point is $1/(1-\eta)$.
For cases of 90\%, 99\%, and 99.9\% efficiency, as one would expect, scaling occurs as $1/N$ for small $j$, and then rapidly depreciates as photon number increases.  This can be explained by the probability distribution being dependant on the detector efficiency, i.e. the losses in the total photon number.  At 99\% for 100 photons one would expect to detect on average 99 photons per measurement, while measuring 10000 photons at an efficiency of 99\% naturally produces much larger gross losses, even though the percentage losses are equivalent for the two cases.  The case of 90\% efficiency is shown in Fig \ref{f:graph90} to illustrate this point.  Important to note is that the average error stays below the SNL, even for 10000 photons.  Between 500 and 10000 one can observe the scaling take on a definite nature of $1/\sqrt{N}$.  The distance below the classical line is within an order of magnitude and is again due to the use of Bayes theorem, coupled to the degree of effective correlation determined by the loss level.  It is also important to note that this line is well below the ``true" SNL, which is given by
\begin{equation}
\Delta \phi = \frac{1}{\sqrt{\eta N}}.	\label{e:class_scale}
\end {equation}
All other efficiencies naturally follow this trend.

\begin{figure}[htb]
\begin{center}
\begin{tabular}{c}
\includegraphics[width= 3in]{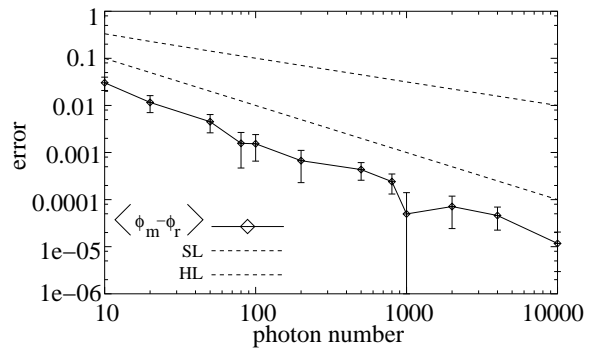}
\end{tabular}
\end{center}
\caption{Average of measured minus real phase for $1-1/N$ efficiency.} \label{f:graphN}
\end{figure}

This leads us to finally consider the particular quantum efficiency $\eta_N$ of Eq.\ \ref{e:eff}. The results of the simulations are plotted in Fig.\ \ref{f:graphN}, which shows, indeed, a behavior comparable to that of the ideal efficiency of Fig.\ \ref{f:graph100} and has been confirmed for phases within the range of $\phi=0$ to several times $\phi=1/N$.

\subsection{Interpretation}

\begin{figure}[htb]
\begin{center}
\begin{tabular}{c}
\includegraphics[width= 3in]{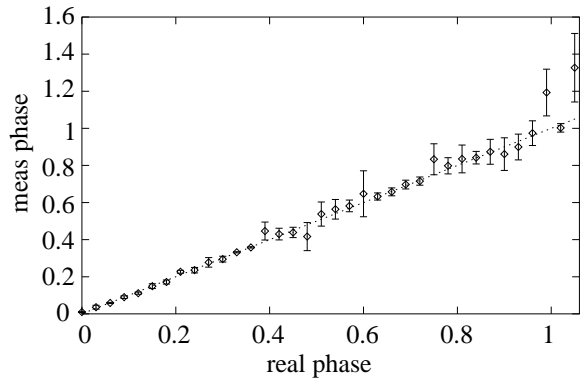}
\end{tabular}
\end{center}
\caption{Average of measured phase vs real phase for 100 photons at 99\% efficiency.} \label{f:graph_phivsphi}
\end{figure}

As the phase angle measured here suggests, in typical experiments the measured phase will need to be within a small range.  As phase angle difference increases one obtains larger error bars on the measured phase, as figure \ref{f:graph_phivsphi} indicates.  
This is true for all efficiencies studied here.  As is clearly evident from the graph, the error bars at higher phase differences are much greater in magnitude than the distance of the measured phase from the line.  The opposite case prevails at low phase differences.  Kim \textit{et al} have suggested that an optimal phase angle exists for each efficiency for which uncertainty is minimized \cite{kim99}.  However, if one opts for the Bayesian postprocessing of data analyzed here, rather than for coincidence measurements, then evidently there is no optimal phase difference, and any minute phase shift around a multiple of $\pi$ radians can be measured with high confidence. 

\section{Conclusion}

We have presented a theoretical study of twin Fock states incident on a Mach-Zehnder interferometer, with simulated measurements based on the Bayesian probability reconstruction first proposed by Holland and Burnett \cite{holland,kim}. The lingering question of the scaling of the phase measurement error with the photon number has been elucidated and it has been found that:

$\bullet$ One may beat the SNL for nonideal detector efficiencies without the signal-to-noise impediment of direct detection of the fourth-order correlation \cite{kim,bouyer}. This paves the way to the realization of a sub-SNL interferometry experiment with phase-locked twin beams \cite{feng3}, an interesting prospect since intensity-difference squeezing is one of the most performant experimental systems available.

$\bullet$ Nonideal efficiencies can still allow one to reach the HL provided that the losses are of the order of $1/N$. Such requirements may be too stringent for photon interferometry, but Bose-Einstein condensates or other atom or trapped-ion experiments may be good candidates.

$\bullet$ The problems found with coincidence detection \cite{kim99} are not relevant to Bayesian twin-mode HLI.

\section{Acknowledgements}

This work is supported by ARO grants DAAD19-01-1-0721 and DAAD19-02-1-0104.

\end{document}